\documentclass[twocolumn,letterpaper,prl,showpacs]{revtex4}

\newcommand{\epl}{Europhys. Lett. }

\newcommand{\pr}{Phys. Rev. }

\newcommand{\jpb}{J. Phys. B }
\newcommand{\UQ}{ARC Centre of Excellence for Quantum-Atom Optics, School of Physical Sciences, University of Queensland, Brisbane, Qld 4072, Australia.}

\usepackage{graphicx}
\usepackage{bm}
\usepackage{amsmath}
\usepackage{color}
\begin{document}

\title{Entanglement properties of degenerate four-wave mixing of matter-waves in a periodic potential}

\author{M.~K. Olsen and M.~J. Davis}

\affiliation{\UQ}

\date{\today}

\begin{abstract}
In a recent experiment Campbell \emph{et al.}\ [\prl {\bf 96}, 020406 (2006)] observed degenerate four-wave mixing of matter-waves in a one-dimensional optical lattice, a process with potential for generating entanglement among atoms. We analyse the essential quantum features of the experiment to show that entanglement is created between the quadratures of the two scattered atomic clouds and is a true many-body (rather than two-body) effect.   
We demonstrate a significant violation of entanglement inequalities that is robust to a moderate level of coherent seeding.  The system is thus a promising candididate for generating macroscopically entangled atomic samples.
\end{abstract}


\maketitle

\textit{Introduction:} A recent development in the field of quantum atom optics has been the proposal of degenerate four-wave mixing of a Bose-Einstein
condensate (BEC) in a periodic potential~\cite{KarenMH}.  The usual quadratic
matter wave dispersion relation in free space ordinarily prevents 
collisional processes in a single condensate generating new momentum components.  However, the dispersion relation of a periodic potential can allow phase-matched two-body collision processes within a single condensate that conserve quasimomentum and energy.  Hence condensates with new momenta can be
spontaneously generated.  The proposal of  Hilligs\o e and M\o lmer~\cite{KarenMH} was
recently implemented experimentally by Campbell \emph{et
al.}~\cite{Wolfgang}, with a BEC loaded into a one-dimensional optical
lattice. When the phase-matching conditions for energy and quasi-momentum were
satisifeld both spontaneous and stimulated scattering were observed, with an
initial state with one quasimomentum being scattered into two different
quasimomentum states.  Recent experimental work by Gemelke \emph{et
al.}~\cite{gemelke} has also investigated phase-matched scattering processes of
matter waves in a driven optical lattice.

Hilligs\o e and M\o lmer~\cite{KarenMH} used the mean-field Gross-Pitaevskii
equation (GPE) to analyse a one-dimensional BEC moving in an optical lattice.
Although multi-mode, the GPE cannot describe spontaneous scattering processes
and their numerical analysis required the initial state to be seeded by hand. 
In addition, a mean-field approach cannot be used to determine the quantum
correlations that are necessary to show entanglement between the scattered
momentum states. 

In this paper we complement the approach of Ref.~\cite{KarenMH} by performing
fully quantum analyses of the dynamics resulting from a simple three-mode
description of the degenerate four-wave scattering process.  While idealised, 
our analysis allows us to demonstrate continuous variable entanglement without the
complications arising from a full multi-mode, multi-dimensional analysis.
Our results suggest that degenerate four-wave mixing could be an efficient way to generate highly entangled atomic samples.  

Campbell \emph{et al.} suggest that ``parametric amplification could also be an efficient means of producing pairs of momentum entangled atoms for quantum information applications.''
In fact, although individual pairs are correlated in the scattering process, they are bosonic and scatter into occupied modes, so that we cannot know which pair a given atom belongs to. It is also difficult to know how individual pairs, if detected, could be used for quantum information processes as this generally requires some other degree of freedom, such as spin~\cite{Pu,Duansqueeze}, which is not present in this experiment.  However, the coherent nature of the scattering allows for the build up of many-body entanglement between the field quadratures, which, while presently difficult to measure experimentally, is robust to losses and seeding. Although other methods have been proposed and demonstrated for the generation of pair-correlated and entangled atoms from BEC~\cite{Pu,Duansqueeze,Vogels,Kherunts1,Haine,Kherunts2}, the method of Refs.~\cite{KarenMH,Wolfgang} appeals because of its relative simplicity. 

\textit{Formalism:}
We consider a condensate loaded into a one dimensional periodic potential in a single Bloch state with quasi-momentum $\hbar k_0$.  For particular combinations of lattice depth and $k_0$, there exists a phase-matched process that conserves both energy and quasi-momentum such that
\begin{equation}
2k_0 = k_1 + k_2,
\qquad 
2\epsilon(k_0) = \epsilon(k_1) + \epsilon(k_2),
\end{equation}
where the generated Bloch modes are $k_1$ and $k_2$, and $\epsilon(k_i)$ is the energy of mode $k_i$. Expanding the full Hamiltonian in terms of Bloch states and using a rotating wave approximation, the interaction picture Hamiltonian is 
\begin{equation}
{\cal H}_{int}=i\hbar\chi\left[\hat{a}_0^{2}\hat{a}_1^{\dag}\hat{a}_2^{\dag}
-\hat{a}_0^{\dag\;2}\hat{a}_1\hat{a}_2\right],
\label{eq:hamiltoniano}
\end{equation}
where $\hat{a}_i$ is the annihilation operator for quasi-momentum mode $k_{i}$, and we have made the transformation $\hat{a}_0 \rightarrow \hat{a}_0 e^{i\pi/4}$. The nonlinear interaction is represented by $\chi$, and is given by
\begin{equation}
\chi = \frac{U_0}{\hbar A_\perp} \int dx \;u_{0}\!(x)^2 u_{1}\!(x) u_{2}\!(x),
\end{equation}
where $U_0 = 4\pi \hbar^2 a/m$, with $a$ the $s$-wave scattering length, $A_\perp$ is the cross-sectional area of the system, and $u_{i}(x)$ is the Bloch state of mode $k_i$ (assumed to be real).

We have made several approximations here. The first is that the Bloch states are a good approximation to the eigenstates of the system.  This will be true as long as the effective interaction strength $\chi$ is sufficiently small.  Using mean-field Bloch states appropriate to the effective potential of the lattice plus initial density will extend the regime of validity for short times while the overall density is unchanged, but will alter the phase matching conditions due to the energy shifts of the Bloch modes.  The three-mode reduction is also only appropriate for short times before scattering into other modes becomes significant. The dimensional reduction is also only appropriate for times short enough that there are no appreciable dynamics in the perpendicular dimensions.

To demonstrate entanglement
we calculate both the Duan criteria~\cite{Duan} (see also Simon~\cite{Simon}) and a set of Einstein-Podolsky-Rosen (EPR) criteria developed by Reid~\cite{eprMDR}, both of which establish the presence of continuous variable bipartite entanglement. These many-body continuous-variable criteria are more appropriate to the present case than any consideration of entanglement between the individual atoms of each scattered pair. This type of many-body entanglement is also more robust to atomic losses~\cite{Kherunts2}.  
For quadrature entanglement, criterion have been outlined
by Dechoum \emph{et al.\/}~\cite{ndturco}, which follow from inequalities
developed by Duan \emph{et al.\/}~\cite{Duan}, based
on the inseparability of the system density matrix, and a
method to demonstrate the EPR paradox~\cite{EPR} using quadratures
was developed by Reid~\cite{eprMDR}. We will briefly outline these criteria here, using the field quadrature operators 
\begin{eqnarray}
\hat{X}_{i} = \hat{a}_i + \hat{a}_i^\dag,
\qquad
\hat{Y}_{i} = -i(\hat{a}_i - \hat{a}_i^\dag).
\end{eqnarray}
To demonstrate entanglement between the modes, we define the combined
quadratures $\hat{X}_{\pm}=\hat{X}_{1}\pm\hat{X}_{2}$ and $\hat{Y}_{\pm}=\hat{Y}_{1}\pm\hat{Y}_{2}$.
Following the treatment of Ref.~\cite{ndturco}, entanglement
is guaranteed provided that 
\begin{equation}
V(\hat{X}_{\pm})+V(\hat{Y}_{\mp})<4.
\label{eq:inequalityduan}
\end{equation}
To examine the utility of the system for the production of states
which exhibit the EPR paradox, we assume that a measurement of the $\hat{X}_{1}$ quadrature, for
example, will allow us to infer, with some error, the value of the
$\hat{X}_{2}$ quadrature, and similarly for the $\hat{Y}_{i}$ quadratures.
By minimising the rms error in these estimates, we find the inferred variances, 
\begin{eqnarray}
V^{inf}(\hat{X}_{1}) &=& V(\hat{X}_{1})-\frac{[V(\hat{X}_{1},\hat{X}_{2})]^{2}}{V(\hat{X}_{2})},\nonumber\\
V^{inf}(\hat{Y}_{1}) &=& V(\hat{Y}_{1})-\frac{[V(\hat{Y}_{1},\hat{Y}_{2})]^{2}}{V(\hat{Y}_{2})},
\label{eq:EPROPA}
\end{eqnarray}
with those for the $k_2$ quadratures being found
by swapping the indices $1$ and $2$. As the $\hat{X}_{i}$ and $\hat{Y}_{i}$
operators do not commute, the products of the actual variances obey a Heisenberg
uncertainty relation, with $V(\hat{X}_{i})V(\hat{Y}_{i})\geq 1$.
Hence we find a demonstration of the EPR paradox whenever 
\begin{equation}
V^{inf}(\hat{X}_{i})V^{inf}(\hat{Y}_{i}) < 1.
\label{eq:demonstration}
\end{equation}

\textit{Analytic results:} 
By analogy to the parametric or undepleted-pump approximation of quantum optics (also sometimes used with coupled atomic and molecular BEC~\cite{Yurovsky}) we can set $\kappa=\chi\langle\hat{a}_{0}^{2}\rangle$ in the Hamiltonian (\ref{eq:hamiltoniano}) with $\kappa$  real, and find the following Heisenberg equations of motion
\begin{eqnarray}
\frac{d\hat{a}_1}{dt} = \kappa\hat{a}_2^{\dag},
\qquad \frac{d\hat{a}_2}{dt} = \kappa\hat{a}_1^{\dag},
\label{eq:heom}
\end{eqnarray}
along with their Hermitian conjugates. The solutions to these equations are well known from quantum optics~\cite{Roy}, and give us 
all the operator moments needed to calculate the entanglement criteria.
With the two modes $k_1$ and $k_2$ initially unpopulated, 
we find
\begin{eqnarray}
V(\hat{X}_{-})+V(\hat{Y}_{+}) &=& 4\left(\cosh 2\kappa t-2\cosh\kappa t\sinh\kappa t\right),\nonumber\\
V^{inf}(\hat{X}_{i})V^{inf}(\hat{Y}_{i}) &=& \frac{1}{\cosh^{2}2\kappa t}\; <\; 1\;\mbox{ for } t > 0,
\label{eq:ancorrelations}
\end{eqnarray}
which obviously violate the appropriate inequalities and can be compared to the numerical results obtained below. 

\begin{figure}[tbhp]
\begin{center}
\includegraphics[width=0.9\columnwidth]{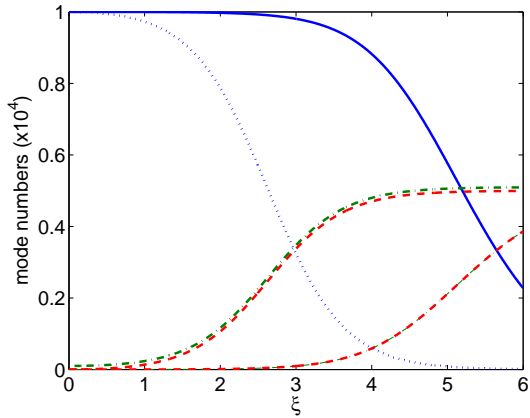}
\end{center}
\caption{The mode occupations as a function of time. The solid line and the lower dashed line are the averages of $4.34\times 10^{6}$ trajectories of the positive-P representation equations for $N_{0}$, $N_{1}$ and $N_{2}$ in the spontaneous case. Note that $N_{1}=N_{2}$ and that the Wigner results are indistinguishable. The dotted line and the upper dashed lines are the Wigner results ($6.6\times 10^{5}$ trajectories), for $N_{0}$, $N_{1}$ and $N_{2}$ with an initial seed, $N_{1}(0)=100$. All quantities plotted in this and subsequent figures are dimensionless.}
\label{fig:numbers}
\end{figure}

\begin{figure}[tbhp]
\begin{center}
\includegraphics[width=0.9\columnwidth]{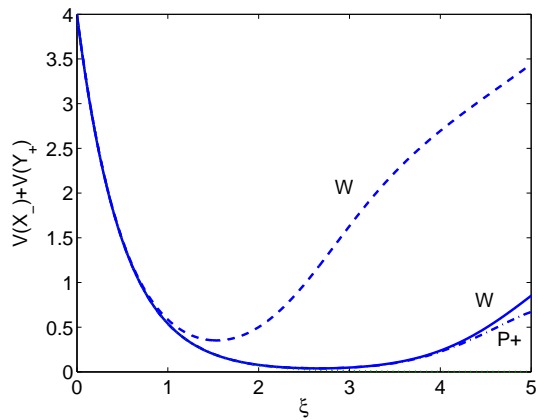}
\end{center}
\caption{The Duan correlation, $V(\hat{X}_{-})+V(\hat{Y}_{+})$, with and without an injected seed. The lower three lines are for the spontaneous case, with the dotted line being analytical, the dash-dotted line being the positive-P prediction, and the full line being the Wigner prediction. The dashed line is the Wigner prediction with $N_{1}(0)=100$.}
\label{fig:Duan}
\end{figure}

\textit{Numerical results: }
The Hamiltonian equations of motion may be mapped exactly onto stochastic differential equations in the positive-P representation~\cite{P+}, following the usual methods~\cite{QNCrispin}. Making the correspondences $\hat{a}_{0}\rightarrow\alpha,\:\:\hat{a}_{1}\rightarrow\beta,\:\:\hat{a}_{2}\rightarrow\gamma$, the stochastic equations are found as
\begin{eqnarray}
\frac{d\alpha}{dt} &=& -2\chi\alpha^{+}\beta\gamma+\sqrt{-\chi\beta\gamma}\;\eta_{1},\nonumber\\
\frac{d\alpha^{+}}{dt} &=& -2\chi\alpha\beta^{+}\gamma^{+}+\sqrt{-\chi\beta^{+}\gamma^{+}}\;\eta_{2},\nonumber\\
\frac{d\beta}{dt} &=& \chi\alpha^{2}\gamma^{+}+\sqrt{\chi\alpha^{2}/2}\left(\eta_{3}+i\eta_{5}\right),\nonumber\\
\frac{d\beta^{+}}{dt} &=& \chi\alpha^{+\;2}\gamma+\sqrt{\chi\alpha^{+\;2}/2}\left(\eta_{4}+i\eta_{6}\right),\nonumber\\ 
\frac{d\gamma}{dt} &=& \chi\alpha^{2}\beta^{+}+\sqrt{\chi\alpha^{2}/2}\left(\eta_{3}-i\eta_{5}\right),\nonumber\\
\frac{d\gamma^{+}}{dt} &=& \chi\alpha^{+\;2}\beta+\sqrt{\chi\alpha^{+\;2}/2}\left(\eta_{4}-i\eta_{6}\right),
\label{eq:Pplus} 
\end{eqnarray}
where the $\eta_{j}$ are real Gaussian noise terms with the correlations
\begin{equation}
\overline{\eta_{j}}=0,
\qquad
\overline{\eta_{j}(t)\eta_{k}(t')}=\delta_{jk}\delta(t-t').
\label{eq:noiseP}
\end{equation}
These equations are integrated over many trajectories, with stochastic averages of the variables becoming equal to normally ordered operator expectation values. 

\begin{figure}[tbhp]
\begin{center}
\includegraphics[width=0.9\columnwidth]{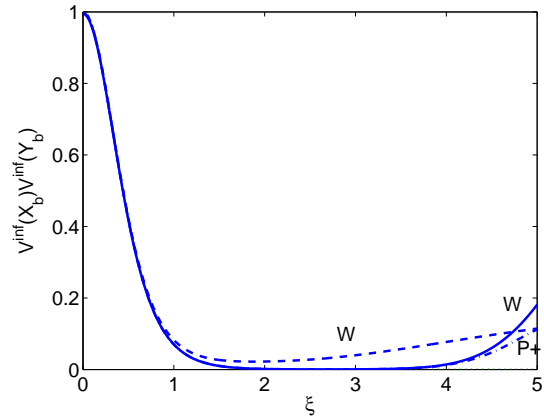}
\end{center}
\caption{The EPR correlation, $V^{inf}(\hat{X}_{1})V^{inf}(\hat{Y}_{1})$. The dotted line is the spontaneous analytical solution, the dash-dotted line is the spontaneous positive-P result, the solid line is the spontaneous Wigner result, and the dashed line is the stimulated Wigner result with $N_{1}(0)=100$.}
\label{fig:EPR}
\end{figure}

Initially we chose $\overline{|\alpha(0)|^{2}}=10^{4}$, with these momentum $k_{0}$ atoms in a coherent state, and $\beta(0)=\gamma(0)=0$.
The numerical solutions are parametrised by $\xi=\chi|\alpha(0)|^{2}t$.  In practice, we found that integration became unstable for times greater than $\xi=6$, and was probably not trustworthy after $\xi\approx 5$. However, this covers the region of maximum violation of the inequalities 
(\ref{eq:inequalityduan},\ref{eq:demonstration}).  

It is also of interest  to compare the positive-P solutions  to those of the
approximate, but stable, truncated Wigner representation.  This has been
used with some success in investigations of BEC~\cite{mjs,
Sinatra2000a,Sinatra2001a,Sinatra2002a,qstate1,qstate2,Fock}, and allows for
the calculation of symmetrically-ordered operator moments.
Again following standard procedures~\cite{QNCrispin}, we can map the system Hamiltonian onto a generalised Fokker-Planck equation for the Wigner pseudoprobability distribution, which has third order derivatives and hence no equivalent stochastic differential equations. Although methods exist for a mapping onto stochastic difference equations~\cite{nossoEPL}, these seldom result in equations which are simple to integrate numerically. We will therefore truncate the third-order terms and map the resulting Fokker-Planck equation onto differential equations for the Wigner variables. This is justified here since the number of particles is much larger than the number of modes. This results in the set of equations
\begin{eqnarray}
\frac{d\alpha}{dt} = -2\chi\alpha^{\ast}\beta\gamma,\quad
\frac{d\beta}{dt} = \chi\alpha^{2}\gamma^{\ast},\quad
\frac{d\gamma}{dt} = \chi\alpha^{2}\beta^{\ast}.
\label{eq:Wminus} 
\end{eqnarray}
Note that, although these equations seem deterministic, the initial variables are chosen from the appropriate Wigner distribution, so that quantum noise is included in the initial conditions. We begin our integrations with coherent states, with $\overline{|\alpha|^{2}}=10^{4}$, $\overline{|\gamma|^{2}}=0$ and  $\beta$ either being vacuum or a small coherent seed of $\overline{|\beta(0)|^{2}}=100$ atoms.

The results are presented in the three figures, which allow us to compare the predictions of the approximate analytic solutions, the formally exact positive-P representation solutions, and those of the truncated Wigner representation. We show the atom numbers in the three modes in Fig.~\ref{fig:numbers} for both the spontaneous and seeded situations.
For the spontaneous case we find that the positive-P and Wigner methods give almost identical results over most of the range shown, thus we can be confident of the Wigner solutions in this regime.  We see that a seed in the $k_{1}$ mode with only $1\%$ of the number of atoms in the $k_{0}$ mode gives appreciably faster scattering into modes $k_{1}$ and $k_{2}$, with almost full occupation at a time when the spontaneous case has seen approximately $10\%$ of the atoms scattered. A similar effect was observed in the experiment of Campbell \emph{et al.}~\cite{Wolfgang}.
As we are interested in the quantum correlations between these two modes, we must also consider the effect of the seeding on these. In the case of the nondegenerate optical parametric amplifier (OPA) with an injected signal, for example, an injection level of $1\%$ had a much less noticeable effect on the mean fields, but was sufficient to almost destroy some quantum correlations~\cite{injectOPA}. In the present case, however, as we can see in Fig.~\ref{fig:Duan} and Fig.~\ref{fig:EPR}, the seeding still allows for considerable violation of the inequalities, especially in the EPR case. We note here that the descriptions are only similar once the parametric approximation is made, which results in a quadratic Hamiltonian for both cases. 
It may be of considerable practical interest that the slightly lesser violations occur for large numbers of atoms in the two scattered modes once the system is seeded, as can be seen by comparison with Fig.~\ref{fig:numbers}. Also of interest is that the EPR product demonstrates entanglement with seeding in a region where the form of the Duan inequality that we have used does not. This is not a contradiction as the most general form of the Duan inequality~\cite{Duan} does not define the quadratures in the symmetric manner we have used here, so that a more appropriate form would be violated.   
  
\textit{Conclusions:} We have calculated the entanglement properties of a Hamiltonian which gives a simplified description of the lattice four-wave mixing experiment of Campbell {\em et al.\/}. We have shown that the system exhibits entanglement between the scattered modes and is a candididate for a demonstration of the EPR paradox with massive particles. Seeding of one of the scattered modes allows for substantially quicker conversion than in the spontaneous case. The entanglement is between the entire modes rather than between individual pairs of scattered atoms and thus is robust to reasonable levels of seeding and is not destroyed by small rates of atomic loss. 
A fully quantum spatial analysis that could more closely describe experiments is currently being investigated.

This research was supported by the Australian Research Council and the Queensland State government.
%

%
%

\end{document}